\documentclass[aps,twocolumn,showpacs,preprintnumbers,prl]{revtex4}
\usepackage{graphicx}
\usepackage{color}
\begin{document}

\title{Theoretical study of ionization of an alkali atom adsorbed on a metal surface by
laser assisted subfemtosecond pulse.}

\author{A.K.Kazansky$^{1,2}$, P.M.Echenique$^{2,3}$}

\affiliation{$^1$ Fock Institute of Physics, The University of St.Petersburg, St.Petersburg,
198504, Russia \\
$^2$ Donostia International Physics Center (DIPC),  Paseo Manuel de Lardizabal, E-20018 San
Sebasti\'an/Donostia, Basque Country, Spain\\
$^3$ Departamento de Fisica de Maeriales and Centro Mixto CCSIG-UPV/EHU, Facultad de Ciencias Quimicas,
UPV/EHU Apdo 1072, 20080 San Sebasti\'an/Donostia, Basque Country, Spain}

\begin{abstract}
The first numerical simulation of the process of ionization of an atom adsorbed on a metal surface by the
subfemtosecond pulse is presented. The streaking scheme is considered, when a weak sub-femtosecond pulse
comes together with a strong IR pulse with a variable delay between them. The problem is analyzed with
numerical solving the non-stationary Schroedinger equation in the cylindrical coordinate.
The results obtained are compared with ones in the gas phase. We show that the surface influences
the DDCS, but the observation of this influence, beside the trivial polarization shift of the energy of the
initial state, requires a quite high experimental resolution.
\end{abstract}

\pacs{79.20Ds, 78.47.jc}
\maketitle


 Development of sub-femtosecond  experimental technique  initiated in 2001 with the paper
 \cite{Hentschel} , providing a tool for monitoring
electrons with sub-atomic resolution in both space and time, creates principally new opportunities for
real-time observation and control of atomic-scale electron dynamics \cite{Cork}. With keeping track on the
temporal evolution of the outgoing electron wavepackets, this technique gives direct time-domain insight into
various aspects of interaction of an electron with very fast relaxing many-electron systems. The attosecond
streaking spectroscopy has already enabled observation of the decay of an inner-shell vacancy through Auger
relaxation in isolated atoms in the gas phase \cite{Auger} and lead to creation of attosecond chronoscopy
\cite{chron}. Extension of the subfemtosecond technique to solids
is of paramount importance. This is because a vast number of transient electron phenomena (e.g., charge
screening,
 collective dynamics in metals and semiconductors, charge transfer in
host–guest systems and other processes linked with electron–electron interaction) at surfaces and in solids
does evolve on a few femtosecond - subfemtosecond time scale.

Although a huge number of papers on attophysics \cite{Ivan} are already published, a very few
proof-of-principle experiments have been performed as yet. The time-dynamics has been observed with an
ingenious 'streaking camera' \cite{streak}, when the IR pulse producing the atto-pulse by the high-harmonics
generation, is simultaneously used for transposing
the electron ejection time into the electron energy spectrum. In parallel with the experimental studies,
development of theory, which should provide full understanding of the experimental findings, is of extreme
importance. In general, the time-independent theoretical paradigm must be supplemented by proliferating
time-dependent theories directly describing attosecond experiments in the time-domain. At the current level
of development of computational technique it becomes possible to build up rather realistic models of the
phenomena with atoms in the gas phase \cite{We1} which are feasible for full, step by step, numerical control
of the results. Meanwhile, comprehensive understanding of dynamic behaviour of electrons in condensed-matter
systems is much more important for promoting experimental investigations and their application in
nanotechnology.
The first experimental extension of the attosecond streaking technique to the processes with the solid
surfaces has been recently performed \cite{Cav}. This is a very important proof-of-principle experiment which
has shown that it is possible to obtain direct time-domain access to charge dynamics at the near-surface
region of metals by probing photoelectron emission from a single-crystal. Theoretical simulations of this
experiment have been presented in \cite{Ech}.

Further development should be directed to investigation of various processes with surface impurities,
including charge transfer, charge screening, image charge creation and decay, etc. The simplest object for
such studies is a single adsorbate on solid surface. In this case, ejection of an electron from an internal
shell of the adsorbate produces abrupt change of  charge distribution external for the solid. Observation of
the energy and angular distribution of the ejected electrons can provide information regarding the relaxation
of the electron system at the surface of the solid. Current state of the computational physics possesses a
potentiality for providing some data on the possible effects, which can be of importance in planning of the
experiments. Here we present the first theoretical simulation of streaking experiment with an atom adsorbed
on a metal surface, restricting ourselves with consideration of the problem within the static one-electron
approximation and do not touch description of the relaxation processes.
We consider a solitary Na adsorbate which is placed at $Z_{ad}=2.5\,a.u.$ from the image plane of Ag(100),
which is at the distance $z_{im}= 2.064\, a.u. $ \cite{Chuklkov} from the outermost layer of the metal toward
the vacuum. The adsorbate is initially ionised to Na$^+$ state. Then it is ionized once more from the
$2p$-shell by the XUV atto-pulse (duration of about 200 as, we consider the pulses with photon energies 50
and 90 eV).
 The ejected electron  moves in the field of the Na$^{++}$ ion $U_2(r)$, in the field of the lattice,
 $U_{s}(z)$, and in the field of a relatively strong IR pulse with $\omega_{IR}=1.6\,eV$,
 which is damped in the metal very efficiently.  The electron ejected into the metal has a rather short mean
 free path (MFP) in the metal ($\lambda_{fr} \approx  10\, a.u.$).
 Both the light pulses are assumed to be polarized perpendicular to the surface.
 In this case, the  component of the angular momentum of the ejected electron
 perpendicular to the surface is a good quantum number. The ionization can proceed from
 both $\sigma$ and $\pi$ states,
 strongly localized in the vicinity of Na nucleus. Studying the electron ejection close to the normal to the
 surface, we presently restrict ourselves with consideration of only the $\sigma$-case.

The basic Schroedinger equation governing the time evolution of the system reads:
\begin{eqnarray}
\imath \frac{\partial}{\partial t} \Psi({\vec r}, t) = ( {\bf H}(\vec r -\vec R_{at})-\imath
\frac{\gamma(z)}{2}-E) \Psi({\vec r}, t) + \nonumber \\(U_{surf}(z) + E_{IR}(t)z)\Psi({\vec r}, t) + \\
\nonumber \frac{1}{2} E_{XUV}(t)z \Phi_{2p}(\vec r-\vec R_{at}) \nonumber
\end{eqnarray}
The Hamiltonian ${\bf H}(\vec r -\vec R_{at})$ describes the interaction of the electron with the $Na^{++}$
core, the corresponding potential has been computed with the Hartry-Slater approximation. The
(pseudo)potential $U_s(z)$ in Eq.(1) mocks the interaction of an electron with the lattice in a finite metal.
For this potential, we use the parametrization by Chulkov et al \cite{Chuklkov} with the parameters for
Ag(100) surface. The damping function $\gamma(z)$ in Eq.(1) takes into account ineleastic electron-electron
collisions; it is non-zero at $z<0$, where $\gamma(z)$ for the electron with energy $E$ in the bulk is
approximated as $\sqrt{2E}/\lambda_f$, with $\lambda_f$ being electron elastic MFP  near the metal surfaces.
It weakly depends on the electron energy, being  close to 5 {\AA} at a few tens eV range \cite{Powell}. We
have set $\lambda_f = 10\,a.u.$

Since the XUV field is rather weak,  the interaction between the active electron and XUV atto-pulse is
described within the rotating wave approximation (RWA, for some details, see \cite{We1}). The quantity $E$ in
(1) is a sum of the electron energy in the initial state, $-|\bar\epsilon_{2p}|$, and the carrier frequency
of the XUV pulse, $\omega_X$ .
 The wave function
of the active electron in the initial state,  $\Phi_{2p}(\vec r-\vec R_{at})$, has been precalculated to take
into account the polarization of this state by the surface potential. The wave function is slightly changed,
but this interaction strongly influences the energy of the initial state: ${\bar \varepsilon}_{2p}= -1.3922
\, a.u.$, while the energy of corresponding state for a single ion $Na^{+}$ within our model is
$\varepsilon_{2p}= -1.6895\, a.u.$ (the experimental value of this energy is $\varepsilon_{2p}^{exp}=
-1.738\, a.u.$.) The shift of the energy of the adsorbate state is close to the classical image shift
$+3/4Z_{ad}=0.3\,a.u.$, for an electron localized on the adsorbate center. The source term $1/2E_{XUV}(t) z
\Phi_{2p}(\vec r-\vec R_{at}) \,$ in Eq.(1) describes ionization of the initial electron state by the XUV
pulse with the envelope $E_{XUV}(t)$. The interaction with the electromagnetic field is taken in the length
form, the factor $1/2$ is due to the RWA used. For the envelope, we have used the form
$\epsilon_X(t)=\exp(-(t-t_{delay})^2/\overline{\tau}_X^2)$
with $\overline{\tau}_X = 125\;as $, FWHM = $\tau_X=0.21\;fs$. The $E_L(z)$ in Eq.(1) describes the NIR laser
electric field:
\begin{eqnarray}
E_L(z) = \left\{
\begin{array}{l l}
  \xi + (z-z_{im}) & \quad \mbox{ $z>z_{im}$}\\
   \xi \exp((z-z_{im})/\xi) + 0.05 & \quad \mbox{ $z<z_{im}$}\\
\end{array} \right.
\end{eqnarray}
with the screening length $\xi = 4\, a.u.$ \cite{Silkin}. The  parameter $z_{im}$ in Eq.(4) is a position of
the image plane. It enters the parametrization of the pseudopotential $U_s(z)$ (see \cite{Chuklkov},
Eq.(2)-(5)), for Ag(100) $z_{im}=2.064\, a.u.$. As in the experiment \cite{Cav}, we assume that the streaking
field incidents on the surface at the Brewster's angle ($\theta_B \approx 77.7^\circ$ for silver). Thus, the
normal to the surface component of the IR field in the metal is about 20 times weaker than the incident
field. For the envelope of the NIR pulse
we use the form
$\epsilon_L(t)=0.5{\cal E}_0 \{1 - \cos[\pi t/\tau_L]\}, \quad 0<t<2\tau_L$
with the FWHM  $\tau_L = 5\;fs$ and $\omega_L = 1.6\, eV,\,\phi=0$ in Eq.(1). The field strength ${\cal E}_0$
is conventionally related to the intensity of the XUV pulse $W$,
we set $W=10^{12}\, W/cm^2$

At variance with the case of a single atom in the gas phase, when computation can be performed with the
spherical coordinate system \cite{We1}, in the present case we are to take into account the strong field by
the nucleus in the cylindrical coordinate system that causes  substantial technical problems. Briefly, the
nonuniform meshes are used in both the $\rho$ and $z$ variables, both the meshes are almost cubic ($\Delta z
\cong z^2$ ) at the center of the adsorbate atom (see \cite{Kazan}). This allows us to take into account the
strong core Coulomb potential with high accuracy. The computations have been performed with the
split-propagation scheme. Strong concentration of the mesh points on the adsorbate core requires some changes
in the Crank-Nicolson algorithm (the computational details will be presented elsewhere). The mesh has covered
uniformly the cylinder $z \in [-80,\,376], \rho \in [0, 450]$ . The time step equals to $0.03\,a.u.$ The
outgoing wave asymptotic condition has been provided by the artificial adsorbing potentials at the edges of
the mesh. Propagation has been performed till $t_{fin}=\tau_{XUV} + 200\,a.u.$. Taking into account that the
electron velocity in vacuum is in 1-2 a.u. range, during this time the essential part of the wave packet
passes through the registering screen (see below), which has been placed at $z_0 = 290 a.u.$ Another problem
in computational scheme with the cylindrical coordinate system is related to a procedure of extraction of
observable quantities from the results of our time-dependent computations. While in the case of a solitary
atom one can completely describe the spectrum of the eigenstates of the ion in a free-field case \cite{We1},
presently construction of such states is not feasible. Here we put forward a different method develops au the
method described in \cite{Sjakste}.

In a far-zone, where only a time-dependent uniform electric field polarized along the $z-$ axis acts on the
electron, the basis functions, normalized at large time over the momentum scale, can be written as:
\begin{eqnarray}
\psi_{\vec q, k_z}( \vec \rho, z, t) = \frac{1}{(2\pi)^{3/2}}\exp \left[ \imath ({\bar Q}(t) + P(t)z + \vec q
\vec \rho )\right].
\end{eqnarray}
The Schroedinger equation
\begin{eqnarray}
\imath \frac{\partial}{\partial t}\psi_{\vec q, k_z}(\vec \rho, z, t) = \Big[ -\frac{1}{2}\Delta + E(t)z
\Big] \psi_{\vec q, k_z}(\vec \rho, z,  t) \nonumber
\end{eqnarray}
leads to the explicit representation for  $Q(t),\; P(t)$:
\begin{eqnarray}
P(t) = k_z + A(t); \quad \quad  A(t)=\int_t^\infty \; d\tau\,E_0(\tau); \nonumber \\
{\bar Q}(t) = -(k_z^2 + q^2) t/2+ k_z Z_1(t) + Z_2(t)/2 ;\\
Z_1(t)= \int_t^\infty \; d\tau\,A(\tau);\quad \quad Z_2(t)=\int_t^\infty \; d\tau\ A^2(\tau) \nonumber
\end{eqnarray}
The basis functions (3) read
\begin{eqnarray}
\psi_{\vec q, k_z}(\vec \rho, z, t) = \frac{1}{(2\pi)^{3/2}}  \exp\left(\imath \Big[-(k_z^2 + q^2)t/2 \right.
 \nonumber \\ \left. + k_z (z+Z_1(t)) +  A(t)z + Z_2(t)/2 + \vec q \vec \rho\ \Big] \right) .
\end{eqnarray}
These functions merge with the conventional plane waves at $t = +\infty $. The expansion of any wave function
\begin{eqnarray}
\Psi(\vec \rho, z, t) = \int_0^{\infty} dk_z d^2\vec q\; f(\vec q, k_z) \psi_{\vec q, k_z}(\vec \rho, z, t),
\end{eqnarray}
which contains only outgoing waves, can be rewritten as
\begin{eqnarray}
\Psi(\vec \rho, z_0-Z_1(t), t) = \nonumber {\rm e}^{\imath \left[A(t)(z_0 - Z_1(t)) +  Z_2(t)/2\right]}\\
\int_0^{\infty}  \frac{d k_z  d^2 \vec q}{(2\pi)^{3/2}}\;  {\rm e}^{\imath \left[-(k_z^2+q^2)t/2 + k_z z_0 +
\vec q \vec \rho \right]}\;f(\vec q, k_z) .
\end{eqnarray}
 Thus, the amplitude $f(\vec q, k_z)$  is the Fourier transform of the wave packet value
 at the moving screen $z(t)=z_0-Z_1(t)$, where $z_0 = z(\infty), \quad \epsilon=(k^2_z+q^2)/2$ :
\begin{eqnarray}
f(k_z, \vec q) = (2 \pi)^{-3/2} k_z {\rm e}^{\imath k_z z_0} \int_{-\infty}^{\infty} \;dt\,{\rm e}^{\imath
[\epsilon t -A(t)z(t) -  Z_2(t)/2 ]}  \nonumber\\
 \int d^2 \vec \rho
 {\rm e}^{- \imath \vec q \vec \rho} \Psi(\vec \rho, z(t),  t) \quad \quad \quad
\end{eqnarray}

Finally, the double differential cross section reads
\begin{eqnarray}
\frac{d^2 \sigma}{d\epsilon d\Omega} = \left[\frac{\sqrt{2\epsilon}}{2 \pi}\right]^3 \cos^2 \theta \;
 \Big|
\int_{-\infty}^{\infty} \;dt\, \nonumber \\ \nonumber{\rm e}^{\imath [\epsilon t -A(t)z(t) -  Z_2(t)/2 ]}
\int d^2 \vec \rho
 {\rm e}^{- \imath \vec q \vec \rho} \Psi(\vec \rho, z(t),  t)   \Big|^2 = \nonumber\\
 \frac{\cos^2 \theta }{2\pi^3 \sqrt{2\epsilon}} \;
 \Big|
\int_{-\infty}^{\infty} \;dt\, {\rm e}^{\imath \epsilon t} \quad \quad \\ \nonumber  \frac{d}{dt} \left\{{\rm
e}^{\imath [ -A(t)z(t) -  Z_2(t)/2 ]} \int d^2 \vec \rho
 {\rm e}^{- \imath \vec q \vec \rho} \Psi(\vec \rho, z(t),  t) \right\}   \Big|^2;
\end{eqnarray}

 Being compared with the method of extraction of information from the time-dependent computations
 used before \cite{We1}, the present approach has a noticeable and universal  advantage. Within the previous
 approach,
we are to perform the computations till the very IR pulse termination and, at the same time, are to guarantee
that the essential part of the wave packet is still in the mesh. This requires using a very large mesh in a
long IR pulse case and, while using the length gauge, this causes at the same time decrease of the time step.
Thus, within the method used previously, the computations for IR pulses longer than 10 fs may  become
unfeasible. Within the present approach, one should keep track on the wave packet propagation only till it
passes the 'screen'. The only restriction on the mesh size is that the moving 'screen', while its
oscillations, should not enter the region where the absorbing potential is operative. Also, the screen should
all the time be in the region where all fields, beside the uniform electric field, can be neglected. This
allows one to deal with rather long IR field. However, we note that the current approach allows to compute
the DDCS  for the electron ejection in a restricted sector in forward and backward directions.

Some results of our computations are given in Fig.1. These computations are performed for a fixed delay
between the center of the XUV atto-pulse and the upset of the IR pulse. For such a delay, the IR field at the
center of the XUV pulse is rather small and therefore the vector-potential reaches its extremum, $A_0 =
-0.128$ a.u. Within the conventional theory of the streaking effect, the energy of the DDCS maximum is equal
approximately to $\epsilon_{max} = \bar {\epsilon}_{2p} + \omega + \sqrt{2 (\bar {\epsilon}_{2p} +
\omega)}A_0 + A_0^2$. The comparison of the maxima positions is  in the table 1. The correspondence is very
good except the case of low frequency $\omega = 45\;eV $, when the whole structure of DDCS is strongly
perturbed with the surface (compare Figs.1a and 1g). Although this frequency range is out of the current
experimental studies \cite{Cork}, it could be of interest because at such low frequencies the ejected
electrons have the velocity comparable with the Fermi velocity in the substrate and therefore in this case
the relaxation of the electron system in the metal after the ejection electron from the internal shell of the
adsorbate could reveal itself most noticeably. Clearly is seen an effect of the conventional polarization
shift in the energy of the initial state due to the interaction of the electron with the image charges
induced in the metal. Currently the experimental studies with atto-pulses are concentrated in the frequency
range about 90 eV. We have obtained a principally noticeable effect of the surface on the DDCE in this
frequency range, but the observation of this influence requires rather high experimental resolution.
\begin{figure}
\includegraphics[width=8.8cm,height=8cm]{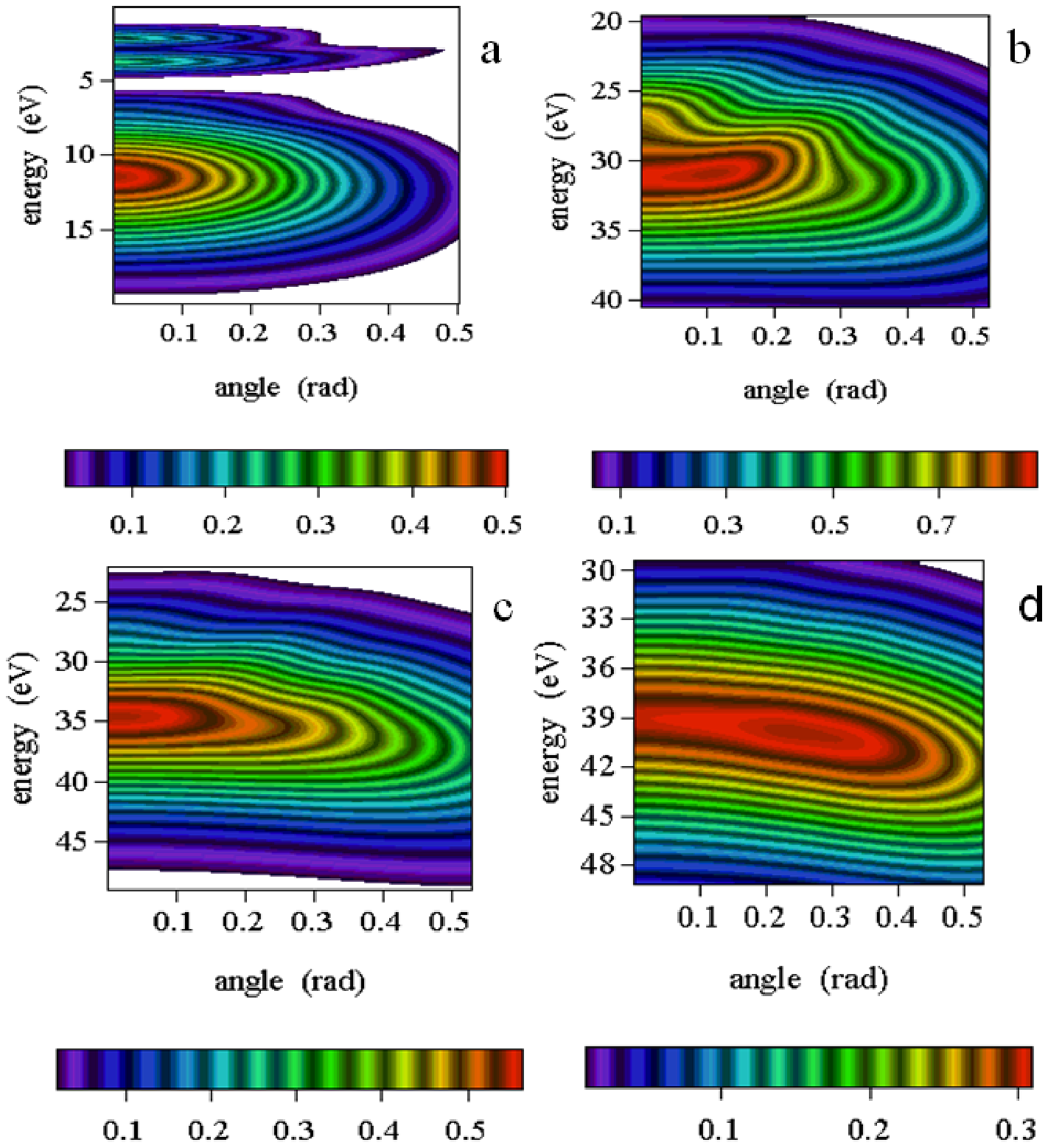}
\includegraphics[width=8.8cm,height=8cm]{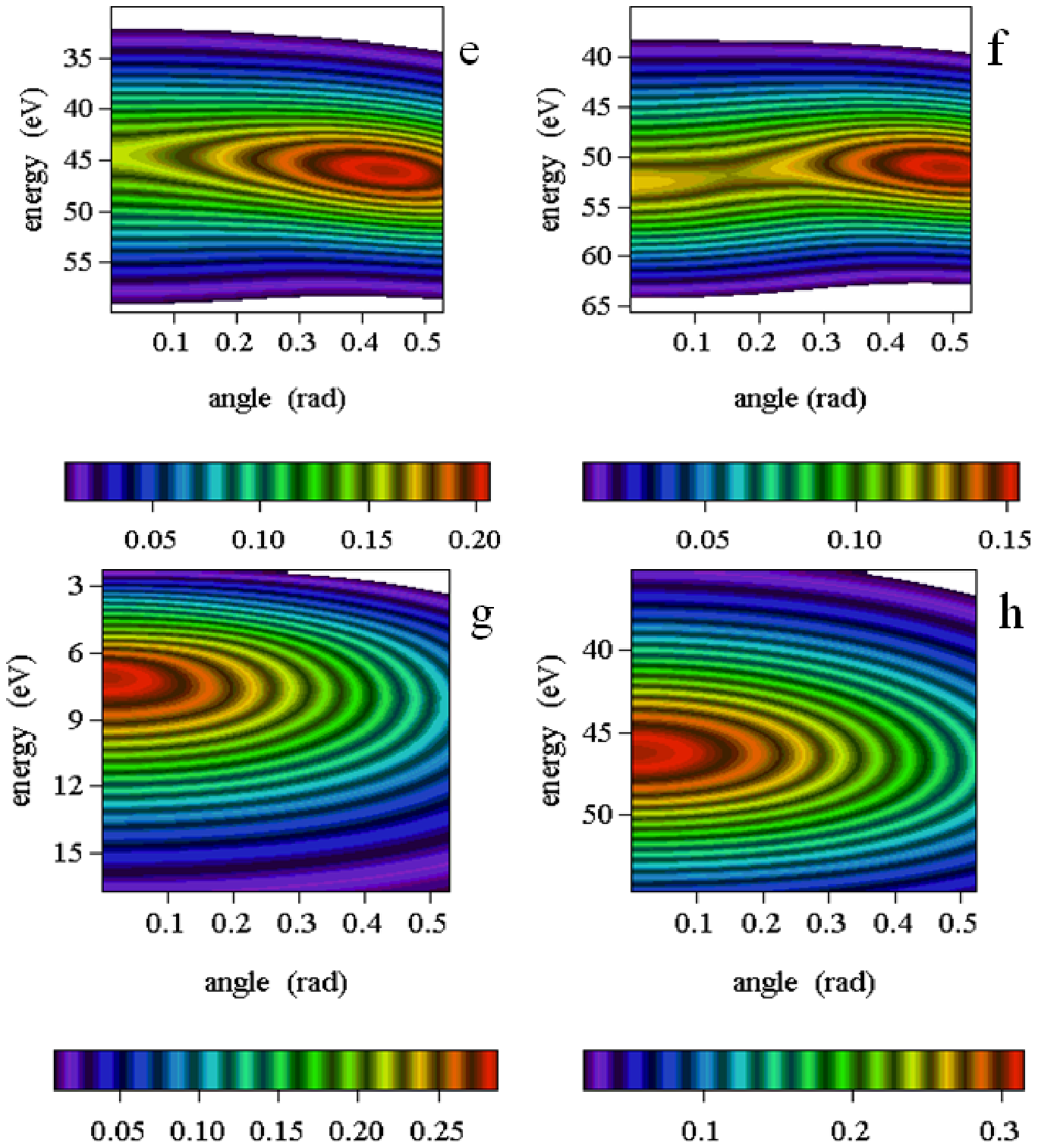}
\caption{ The results of computations of DDCS $d^2 \sigma /d\epsilon d\Omega$ as functions of the final
electron energy ($\epsilon$) and angle between the final electron momentum and the normal of the surface
($\theta$) for different frequencies of the XUV pulse ($\omega_{XUV}$). In Figs.a-f  the DDCS for ionization
of
the adsorbate are given, in Figs. g,h the results for ionization of a free Na$^+$ ion.\\
a: $\omega_{XUV}= 45\;eV$; b: $\omega_{XUV}= 75\;eV$; c: $\omega_{XUV}= 80\;eV$; d: $\omega_{XUV}= 85\;eV$;
e: $\omega_{XUV}= 90\;eV$; f: $\omega_{XUV}= 95\;eV$; g: $\omega_{XUV}= 53.1\;eV$; h: $\omega_{XUV}=
98.1\;eV$} \label{fig:l}
\end{figure}
\begin{table}
\caption{Comparison of the energies (in eV) of maxima DDSC obtained in computations $\epsilon_{max}$ and
$\epsilon_{eff}$ obtained within the conventional theory of streaking. }
\begin{tabular}{|p{0.9cm}||p{0.9cm}|p{0.9cm}|p{0.9cm}|p{0.9cm}|p{0.9cm}|p{0.9cm}|p{0.9cm}| p{0.9cm}|}
\hline
$\omega$ & 75 & 80 & 85 & 90 & 95 & 53.1 & 98.1\\
\hline
$\epsilon_{max}$ & 31 & 35 & 41 & 46 & 51 & 5 & 46\\
\hline
$\epsilon_{eff}$ & 31.6 & 36.2 & 40.9 & 45.5 & 50.2 & 4.82 & 45.5\\
\hline
\end{tabular}
\end{table}

In conclusion, a method for numerical simulation of DDCS of electron ejection from an internal shell of an
atom adsorbed on a metal surface by attosecond XUV pulse accompanied with strong IR laser field is developed.
It is shown that the surface produces a noticeable effect in DDCS. The method can be applied for numerical
simulation of the effects of relaxation in the electron system of substrate caused by the abrupt change of
the external charge distribution due to absorption of an attosecond pulse.

AKK deeply acknowledges financial support from the Ikerbasque Foundation. PEM acknowledges partial support
from the University of the Basque Country (9/UPV 00206.215-13639/2001), the Basque Unibersitate eta Ikerketa
Saila and the Spanish Ministerio de Education y Ciencia (MEC) (FIS 2004- 06490-C03-01 and CSD2006-53). We are
thankful to F Krausz, A G Borisov, and N M Kabachnik for useful discussions.


\begin{thebibliography}
{99}
\bibitem{Hentschel}
   M.Hentschel {\it et al}. Nature 414, 509 (2001)
\bibitem{Cork}
   P.B.Corkum and F. Krausz.  Nature Physics , VOL 3 , 381,  2007\\
\bibitem{Auger}
     Drescher M,
  {\em et al}, Nature,  {\bf 419}, 803  (2002) \\
Drescher M, {\it et al}
J.Electron Spectrosc.Relat.Phenom., {\bf 137-140}, 259 (2004) \\
\bibitem{chron}
M. Uiberacker {\it et al},  Nature {\bf 446} 627, (2007)\\
Th. Uphues {\it et al}, New J. Phys. {\bf 10}, 025009 (2009)\\
A.K. Kazansky and N.M. Kabachnik, J. Phys. B, {\bf 41}, 135601 (2008)\\
\bibitem{Ivan}
F.Krausz and M.Ivanov. Rev. Mod. Phys, {\bf 81}, 163, 2009.\\
 \bibitem{streak}
  Itatani J.,
  {\em et al}, Phys. Rev. Lett.,  {\bf 88}, 173903 2002 \\
    Kitzler M.,
 {\em et al}, Phys.Rev.Lett, {\bf 88}, 173904 \\
\bibitem{We1}
 A.K.Kazansky, N.M.Kabachnik. J.Phys.B, {\bf 40}, 2163, (2007);
  {\it ibid} {\bf 39}, L53 (2006);
  {\it ibid} {\bf 40}, 3413 (2007)\\
 \bibitem{Cav} Cavalieri AL
{\em et al}, Nature, {\bf 449} 1029-1032 2007
\\
\bibitem{Ech}
A.K.Kazansky and P.M.Echenique. Phys.Rev.Lett. {\bf 102} 177401 (2009) \\
C.-H. Zhang and U. Thumm, Phys. Rev. Lett. {\bf 102}, 123601 (2009).
\bibitem{Chuklkov} E.V.Chulkov, V.M.Silkin, P.M.Echenique. Surf.Sci.,  {\bf 437}, 330, 1999\\
\bibitem{Powell} C. J. Powell and A. Jablonski, J. Phys. Chem. Ref. Data
28, 19 (1999)\\
\bibitem{Silkin} V.M.Silkin, private communication.\\
\bibitem{Kazan}
A.K.Kazansky, J.Phys.B {\bf 31}, L579 (1998)\\
A.K.Kazansky, A.G.Borisov, and J.-P.Gauyacq, Nucl.Instrum.Methods Phys.Res. B, {\bf 137}, 21 (1999)\\
\bibitem{Sjakste}
J. Sjakste {\it et al},\mathfrak{} J. Phys. B {\bf 37}, 1593-1603 (2004)
\end {thebibliography}
\end{document}